\input harvmac.tex
\input labeldefs.tmp
\writedefs
\overfullrule=0mm
\hfuzz 10pt

\input amssym.def
\input amssym.tex

\def\modif#1{} 

\input epsf.tex
\newcount\figno
\figno=0
\def\fig#1#2#3{
\par\begingroup\parindent=0pt\leftskip=1cm\rightskip=1cm\parindent=0pt
\baselineskip=11pt
\global\advance\figno by 1
\midinsert
\epsfxsize=#3
\centerline{#2}
\vskip 10pt
{\bf Fig. \the\figno:} #1\par
\endinsert\endgroup\par
}
\def\figlabel#1{\xdef#1{\the\figno}\writedef{#1\leftbracket \the\figno}}
\def\encadremath#1{\vbox{\hrule\hbox{\vrule\kern8pt\vbox{\kern8pt
\hbox{$\displaystyle #1$}\kern8pt}
\kern8pt\vrule}\hrule}}
\newdimen\xfigunit
\global\xfigunit=4144sp
%
\newcount\x \newcount\y
\newcount\newmag\newcount\oldmag
\def\pic(#1,#2)(#3,#4)#5{%
\global\x=-#3%
\global\advance\x by-#1%
\global\y=-#4%
\global\newmag=\epsfxsize%
\global\divide\newmag by \xfigunit%
\def\epsfsize##1##2{\expandafter\epsfxsize%
\global\oldmag=##1\global\divide\oldmag by \xfigunit}%
\epsfbox{#5}%
}
\def\put(#1,#2)#3{%
{%
\advance\x by#1\advance\y by#2%
\multiply\x by \newmag%
\divide\x by \oldmag%
\multiply\y by \newmag%
\divide\y by \oldmag%
\rlap{\kern\x\xfigunit%
\raise\y\xfigunit\hbox{#3}}}}
%
%
%
%
%
\def\frac#1#2{{\scriptstyle{#1 \over #2}}}

%
%

%
\def\({ \left( }\def\[{ \left[ }
\def\){ \right) }\def\]{ \right] }
%


\def\IR{\relax{\rm I\kern-.18em R}}
\font\cmss=cmss10 \font\cmsss=cmss10 at 7pt
\def\IZ{\relax\ifmmode\mathchoice
{\hbox{\cmss Z\kern-.4em Z}}{\hbox{\cmss Z\kern-.4em Z}}
{\lower.9pt\hbox{\cmsss Z\kern-.4em Z}}
{\lower1.2pt\hbox{\cmsss Z\kern-.4em Z}}\else{\cmss Z\kern-.4em Z}\fi}
\def\inbar{\,\vrule height1.5ex width.4pt depth0pt}
\def\IB{\relax{\rm I\kern-.18em B}}
\def\IC{\relax\hbox{$\inbar\kern-.3em{\rm C}$}}
\def\ID{\relax{\rm I\kern-.18em D}}
\def\IE{\relax{\rm I\kern-.18em E}}
\def\IF{\relax{\rm I\kern-.18em F}}
\def\IG{\relax\hbox{$\inbar\kern-.3em{\rm G}$}}
\def\IH{\relax{\rm I\kern-.18em H}}
\def\II{\relax{\rm I\kern-.18em I}}
\def\IK{\relax{\rm I\kern-.18em K}}
\def\IL{\relax{\rm I\kern-.18em L}}
\def\IM{\relax{\rm I\kern-.18em M}}
\def\IN{\relax{\rm I\kern-.18em N}}
\def\IO{\relax\hbox{$\inbar\kern-.3em{\rm O}$}}
\def\IP{\relax{\rm I\kern-.18em P}}
\def\IQ{\relax\hbox{$\inbar\kern-.3em{\rm Q}$}}
\def\IGa{\relax\hbox{${\rm I}\kern-.18em\Gamma$}}
\def\IPi{\relax\hbox{${\rm I}\kern-.18em\Pi$}}
\def\ITh{\relax\hbox{$\inbar\kern-.3em\Theta$}}
\def\IOm{\relax\hbox{$\inbar\kern-3.00pt\Omega$}}


\def\oh{{1\over 2}}


\def\dim{{\rm dim\,}}\def\mod{{\rm mod\,}}

\def\nind{\noindent}

\def\ie{{i.e.\ }}
\def\\#1 {{\tt\char'134#1} }

\catcode`\@=11
\def\Eqalign#1{\null\,\vcenter{\openup\jot\m@th\ialign{
\strut\hfil$\displaystyle{##}$&$\displaystyle{{}##}$\hfil
&&\qquad\strut\hfil$\displaystyle{##}$&$\displaystyle{{}##}$
\hfil\crcr#1\crcr}}\,}   \catcode`\@=12
\def\encadre#1{\vbox{\hrule\hbox{\vrule\kern8pt\vbox{\kern8pt#1\kern8pt}
\kern8pt\vrule}\hrule}}
\def\encadremath#1{\vbox{\hrule\hbox{\vrule\kern8pt\vbox{\kern8pt
\hbox{$\displaystyle #1$}\kern8pt}
\kern8pt\vrule}\hrule}}


\def\tvp{\vrule height 2pt depth 1pt} 
\def\thp{\vrule height 0.4pt width 0.35em}
\def\cc#1{\hfill#1\hfill}

\setbox101=\vbox{\offinterlineskip
\cleartabs                              
\+ \thp&\cr
\+ \tvp\cc{}&\tvp\cr 
\+ \thp&\cr  }

\def\bY{\bar Y}
\def\ommit#1{}
  
\def\mathCO#1{{\tt math.CO/#1}}  \def\condmat#1{{\tt cond-mat/#1}}

%
%

\def\JPhA#1{{\it J. Phys. A} {\bf #1}}

\lref\MRR{W.H. Mills, D.P. Robbins and H. Rumsey, 
{\sl Alternating sign matrices and descending plane partitions},
{\it J. Combin. Theory} Ser.A, {\bf 34} (1983) 340-359.}
\lref\ZK{D. Zeilberger, {\sl Proof of the alternating sign matrix
conjecture}, {\it Electr. J. Combin.} {\bf 7} (2000) R37\semi
G. Kuperberg, {\sl Another proof of the alternating sign matrix
conjecture}, {\it Int. Math. Res. Notes} (1996) 139-150, \mathCO{9712207}. }

\lref\Pro{J. Propp, {\sl The many faces of alternating-sign matrices}, \mathCO{0208125}.}
\lref\Bress{D. Bressoud, {\sl Proofs and Confirmations: The Story of
the Alternating Sign Matrix Conjecture}, Cambridge Univ. Pr., 1999.}
\lref\RS{A.V. Razumov and Yu.G. Stroganov, 
{\sl Spin chains and combinatorics}, 
\JPhA{34} (2001) 5335-5340, \condmat{0012141};  
{\sl Combinatorial nature
of ground state vector of O(1) loop model}, \mathCO{0104216}.}
\lref\BdGN{M.T. Batchelor, J. de Gier and B.Nienhuis,
{\sl The quantum symmetric XXZ chain at $\Delta=-{1\over 2}$,
alternating sign matrices and plane partitions}, 
\JPhA{34} (2001) L265-270, \condmat{0101385}.}
\lref\Ku{G. Kuperberg, {\sl Symmetry classes of alternating-sign
matrices under one roof}, \mathCO{0008184}.}
\lref\Rob{D.P. Robbins, {\sl Symmetry classes of alternating-sign
matrices},  \mathCO{0008045}.}
\lref\Stoc{P.K. Stockmeyer, {\sl The charm bracelet problem and its 
applications}, {\it Lect. Notes Math.} {\bf 406} (1974) 339-349.}
\lref\Wie{B. Wieland, {\sl  A large dihedral symmetry of the set of
alternating-sign matrices}, 
 {\it Electron. J. Combin.} {\bf 7} (2000) R37, 
\mathCO{0006234}.}

\lref\PRdGN{P. Pearce, V. Rittenberg, J. de Gier and B. Nienhuis, 
{\sl Temperley-Lieb stochastic processes}, \JPhA{35} (2002) L661-668.}
\lref\PDF{ P. Di Francesco, private communication.}
\lref\NAM{ Nguyen Anh-Minh, private communication.}

\lref\dGNPR{J. de Gier, B. Nienhuis, P. Pearce and V. Rittenberg, 
{\sl The raise and peel model of a fluctuating interface},
\condmat{0301430}.}

\lref\Stan{R.P. Stanley, {\sl Enumerative Combinatorics}, vol. 2
chap. 7.21, Cambridge Univ. Pr.}
\lref\vLW{J.H. van Lint and R.M. Wilson, {\sl A Course in
Combinatorics},  Cambridge Univ. Pr., 1992.}
\lref\DW{D. Wilson,  private communication.}

\lref\MNdGB{S. Mitra, B. Nienhuis, J. de Gier and M.T. Batchelor,
{\sl Exact expressions for correlations in the ground state 
of the dense $O(1)$ loop model}, to appear.}
\Title{
}
{{\vbox {
\vskip-10mm
\centerline{ On the Counting of Fully Packed Loop Configurations}
\medskip\centerline{Some new conjectures }
}}}
\medskip
\medskip
\centerline{J.-B. Zuber}\medskip
\centerline{\it C.E.A.-Saclay, Service de Physique Th\'eorique de Saclay,}
\centerline{\it CEA/DSM/SPhT, Unit\'e de recherche associ\'ee au CNRS}
\centerline{\it F-91191 Gif sur Yvette Cedex, France}

\vskip .2in

\noindent 
New conjectures are proposed on the numbers of FPL configurations
pertaining to certain types of link patterns. 
Making use of the Razumov and Stroganov Ansatz, 
these conjectures are based on the analysis  
of the ground state of
the Temperley-Lieb chain, for periodic boundary conditions 
and so-called ``identified connectivities'', up to size $2n=22$.

\bigskip

\Date{9/03} 
%
\vfill\eject

\newsec{Introduction }
\fig{The $n\times n$ grid (here $n=3$ and $n=4$) with $2n$ external links 
occupied}{\epsfbox{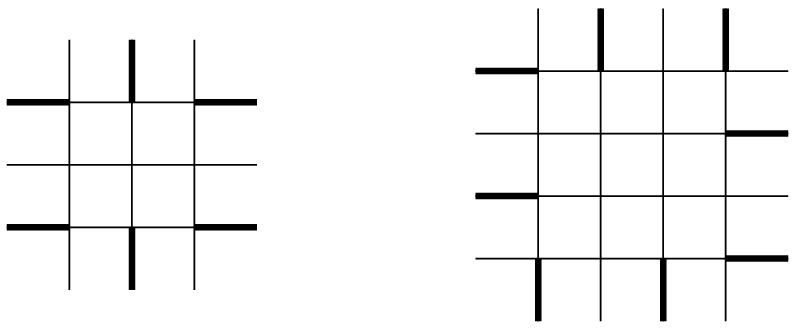}}{6cm}\figlabel\fplg

\noindent
Consider a $n\times n$ square grid, with its $4n$ external links,
see Figure \fplg. We are interested in {\it Fully Packed Loops} (FPL 
in short), {\ie} sets of disconnected paths
which pass through each of the $n^2$ vertices of the grid
and exit through $2n$ of the external links, every second of them
being occupied (see figure \fplqb\ for the case $n=4$).

\fig{The  42 FPL configurations on a $4\times 4$ grid. Configurations
corresponding to distinct link patterns are separated by semi-colons.  }
{\epsfbox{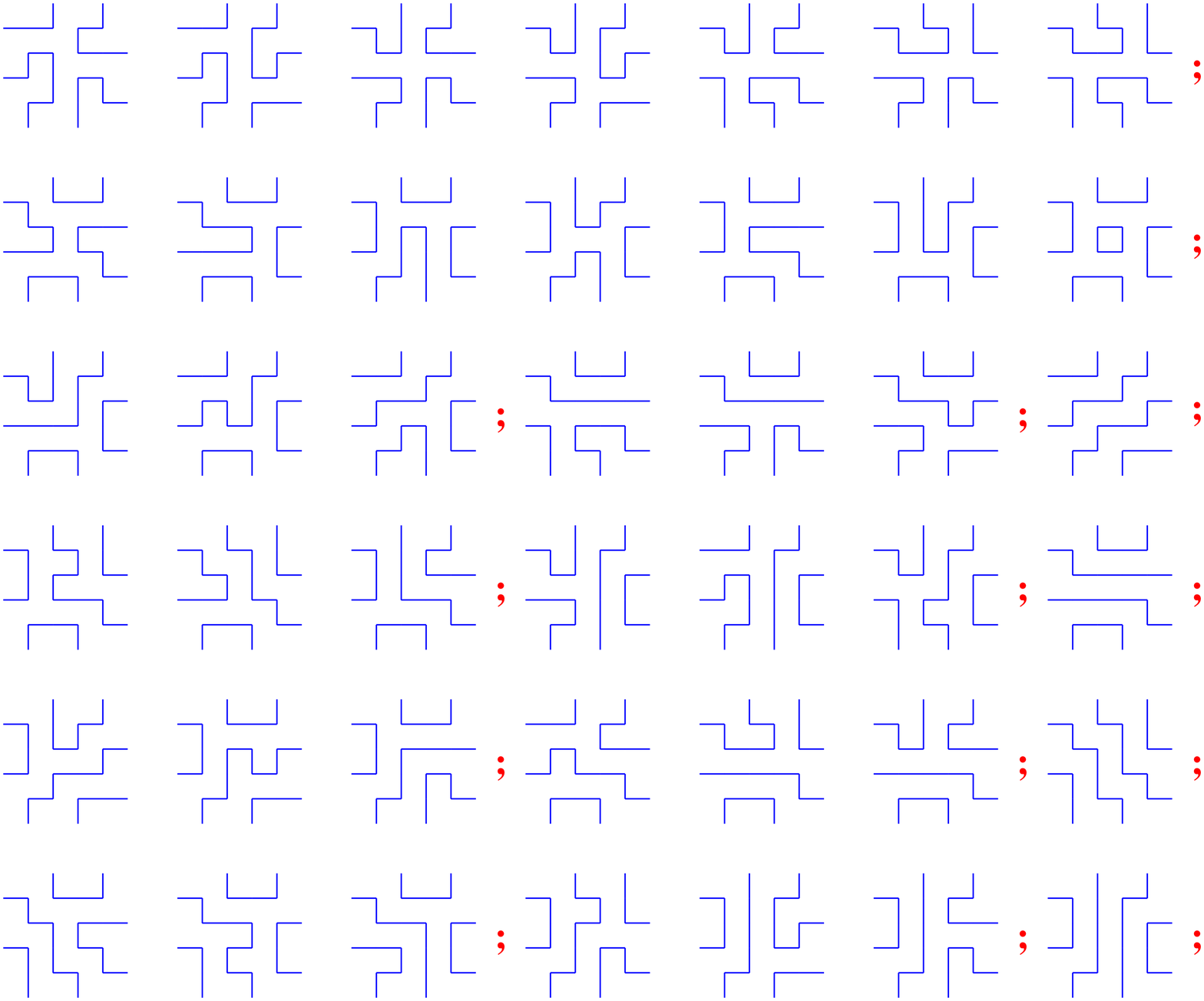}}{10cm}\figlabel\fplqb
There is a simple one-to-one correspondence between 
such FPL and alternating-sign matrices (ASM), obtained as 
follows: divide the $n^2$ vertices into odd and even as usual, and 
 associate  $+1$ (resp. $-1$) to each horizontal
segment of the path passing through an even (resp. odd) vertex, 
the opposite if the
segment is vertical, and $0$ if the path has a corner at that vertex.
\fig{FPL--ASM correspondence}{\epsfbox{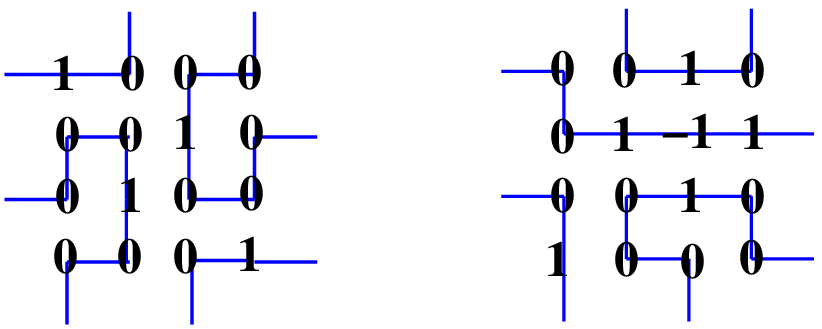}}{6cm}\figlabel\fplasm
 This prescription associates an $n\times n$ ASM 
matrix to the FPL configuration 
in a one-to-one way. Thanks to the celebrated result on ASM's \refs{\MRR,\ZK}, 
the total number of FPL is thus known  to be 
\eqn\ASM{
A_n= \prod_{j=1}^n {(3j-2)!\over (n+j-1)!}\ .}
For a review, see \refs{\Bress,\Pro}. 

Considering FPL rather than ASM enables one to ask different
questions, which are more natural in the path picture.  
Each FPL configuration defines 
a certain connectivity pattern, or {\it link pattern}, between the 
$2n$ occupied external links.  
Let $A_n(\pi)$ be the number of FPL configurations for a given 
link pattern $\pi$. We want to collect results and conjectures
about these numbers $A_n(\pi)$. The next two sections recall
well-known results and conjectures, 
while the following one gathers a certain number
of conjectures which had not appeared in print before to the best
of my knowledge. It is hoped that they will stimulate someone else's
interest or suggest to an ingenious reader 
a connection with a different problem.


\newsec{Counting the orbits}

\fig{The three  link patterns up to rotations and reflections 
for $n=4$}{\epsfbox{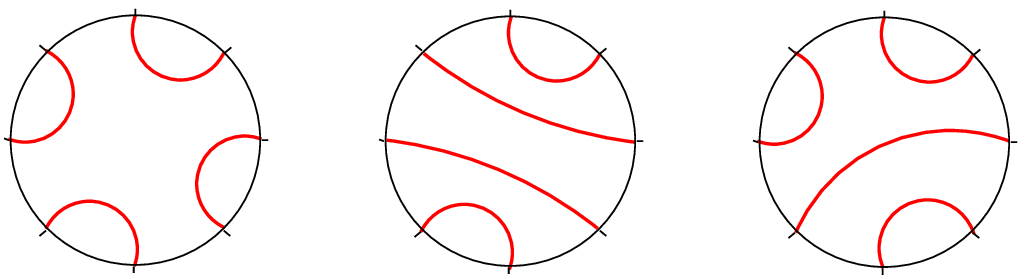}}{6cm}\figlabel\linkp

\noindent 
Although the problem of evaluating the $A_n(\pi)$ seems to admit 
only the symmetries of the 
square, it is convenient to represent the link patterns by 
arches connecting $2n$ points regularly distributed on a circle
(see figure \linkp).

Wieland \Wie\ has proved the remarkable result that $A_n(\pi)$ depends
only on the equivalence class of $\pi$ under the action of the dihedral group 
$D_n$ generated by the
rotations by $2 \pi /2n$ and reflections across any diameter 
passing through a pair of these points. 
While it is easy to convince oneself that the number of
link patterns equals
$ C_n ={(2n)!\over n!(n+1)!} $ (the Catalan number), computing  
the number $O_n$ of orbits under the action of $D_n$, i.e. of 
independent link patterns, is more subtle and appeals to 
Polya's theory of orbit counting  (see for example \vLW). 
In fact,  using an alternative
representation by the dual graph (see Figure 5), 
one realizes that these orbits 
are in one-to-one correspondence with the {\it projective planar
trees} (PPT's) on $n+1$ points, 
whose generating function $T(x)=\sum_{n=1} O_n x^n$ has been computed by
Stockmeyer \Stoc. We recall here his result for the convenience of 
the reader. 
Let $z_1,z_2,\cdots,z_n$ and $y$ be $n+1$ indeterminates and define
the {\it modified cycle index} of the dihedral group $D_n$ as
\eqn\Zindex{Z(D_n^*;z_1,z_2,\cdots,z_n,y)=
{1\over 2n}\sum_{i|n}\phi(i) z_i^{n/i}
+\cases{\oh y z_2^{(n-1)/2} & if $n$ is odd \cr
{1\over 4} y^2z_2^{(n-2)/2}+z_2^{n/2} & if $n$ is even, \cr}}
where $\phi(n)$ is the Euler totient function, counting the number of 
positive integers less than $n$ which are relatively prime to $n$.
Let $c(x)=\sum_{n=0} {(2n)!\over n!(n+1)!} x^{n+1}$ be the generating
function of the Catalan numbers and define 
  $a(x)=x/(1-x-c(x^2))$. The generating function $R(x)$ of the numbers
of rooted planar projective trees is then given by 
\eqn\Rfn{R(x)= x Z(D_n^*;c(x),c(x^2),\cdots,c(x^n),a(x))}
while the one of unrooted PPT's, which we want, is
\eqn\Tfn{T(x)=R(x)-Z(D_2^*;c(x),c(x^2),a(x))+c(x^2)\ .}
One finds
\eqn\genfn{ 
 T(x)=x+x^2+2x^3+3x^4+6x^5+12x^6+27x^7+65x^8+175x^9+490x^{10}+1473x^{11}+4588x^{12}+\cdots}

\fig{The dual picture of a link pattern as a planar 
tree}{\epsfbox{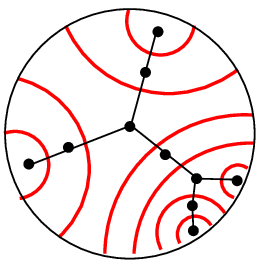}}{2cm}\figlabel\pptree

In Table 1, we list the values of $A_n$,  
$C_n$ and $O_n$ for low values of $n$.
 
\bigskip
\hglue-15mm\vbox{\halign{#\hfill:\quad && \hfill#\hfill\quad\cr
$n$& 1 & 2 & 3 & 4 & 5 & 6 & 7 & 8 & 9 & 10  & 11\cr
$\scriptstyle{A_n}$ &$\scriptstyle{1}$ &$\scriptstyle{ 2}$ &$\scriptstyle{ 7}$ &$\scriptstyle{ 42}$ &$\scriptstyle{ 429}$ &$\scriptstyle{7436}$ & $\scriptstyle{218348}$ &
$\scriptstyle{10850216} $ &$\scriptstyle{911835460} $ &$\scriptstyle{129534272700} $ &
$\scriptstyle{31095744852375}$
\cr
$C_n$& 1 &  2 & 5 & 14 & 42 &  132 & 429 & 1430 & 4862 & 16796 & 58786  \cr
$O_n$ & 1 & 1 & 2 &  3 & 6 & 12 & 27 & 65 & 175 & 490 &1473  \cr
}}
In the following, we use either the notation of link patterns with 
arches, or their dual PPT graphs, or both. 
The $2n$ external links are numbered from $1$ to $2n$ in cyclic
order. A link pattern $\pi_a$ may be regarded as an involutive
permutation on $\{1,\cdots,2n\}$, with $\pi_a(i)=j$ for each arch 
connecting $i$ and $j$. 


\newsec{The $A_n(\pi)$ as solutions of a linear problem}
\nind
The work of Razumov and Stroganov \RS\ and Batchelor, de Gier
and Nienhuis~\BdGN\ contains a certain number of conjectures on the 
numbers $A_n(\pi)$. The most remarkable one connects them to a
linear problem, as follows. 

The periodic Temperley-Lieb algebra $PTL_{p}(\beta)$ is the algebra 
generated by the identity and $p$ generators $e_i$, with the index $i$
running on $\{1,\cdots, p\}$ modulo $p$, satisfying 
\footnote{$^{(1)}$}{Note that because we are working on a disk rather than a cylinder
(more precisely we let the $e$'s act on link patterns on a disk), 
we don't have to  consider non-contractible loops
nor to introduce additional relations between the $e$'s: 
we are working in the so-called
``identified connectivities'' periodic sector \PRdGN.}
\eqnn\TLA
$$\eqalignno{e_i^2&=\beta e_i 
\cr
e_i e_{i\pm 1}e_i&=e_i & \TLA \cr
e_i e_j&=e_j e_i \qquad {\rm if \ \ }|i-j|\ \mod p >1 \ .\cr
}$$ 
There exists a faithful graphical representation of $PTL_p$, see 
figure. 
\fig{The graphical representation of the Temperley-Lieb algebra $PTL_p$,
with $i=1,\cdots, p$ mod $p$.}{\epsfbox{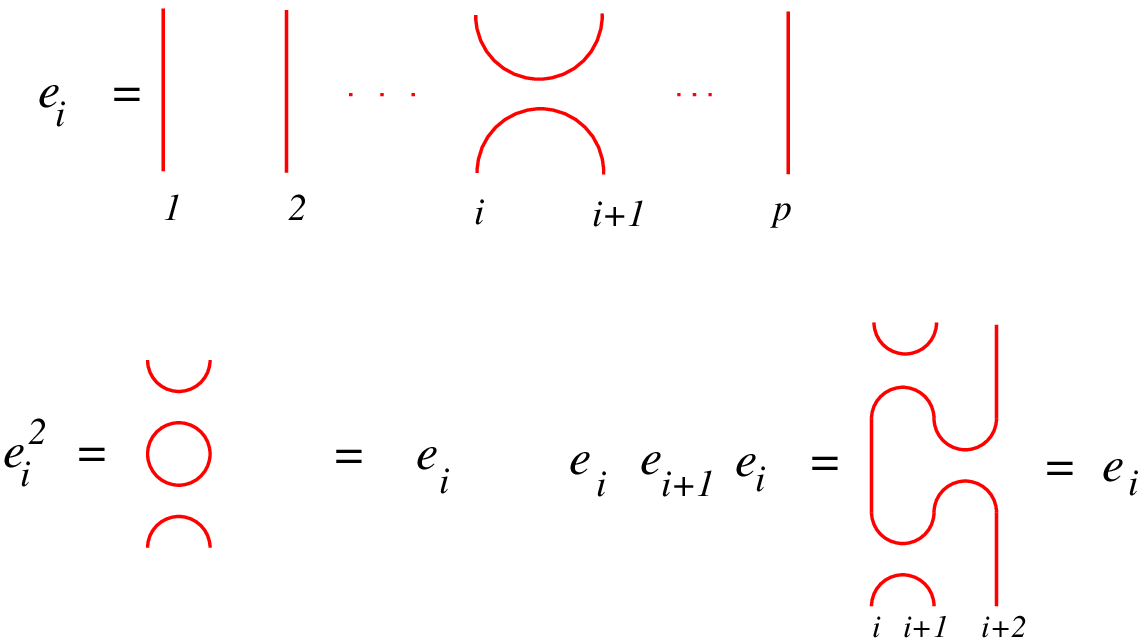}}{6cm}\figlabel\TLa

Now take $\beta=1$ and let $PTL_{2n}(1)$ act on the link patterns
$\pi_a$, $a=1,\cdots, C_n$: using the graphical representation above, 
it is clear that
$e_i$ maps $\pi_a$ on itself if $\pi_a(i)=i+1$, while $\pi_b=e_i \pi_a$ 
connects 
 $j$ and $k$ (as well as $i$ and $i+1$) 
if $\pi_a(i)=j$, $\pi_a(i+1)=k$. Define  
\eqn\Hamil{H=\sum_{i=1}^{2n} e_i\ .} 
In the basis $\{\pi_a\}$, 
$H$ admits $(1,1,\cdots,1)$ 
as   a left eigenvector of eigenvalue $2n$. This is
its largest eigenvalue, and as the matrix $H$ is irreducible and 
has non negative entries, one may use Perron-Frobenius theorem to assert
that the right eigenvector for that largest eigenvalue 
must have non negative components. According to \RS, one has \par
\noindent{\bf Conjecture 1.} \RS\ 
{\sl The right eigenvector of $H$ of eigenvalue 
$2n$ is $\Psi=\sum_a A_n(\pi_a) \pi_a$}
\eqn\eigenv{\sum_{i=1}^{2n} e_i \sum_a A_n(\pi_a) \pi_a = 2n
\sum_a A_n(\pi_a) \pi_a\ .}

\fig{ the configurations of (a) smallest, (b) largest 
component}{\epsfbox{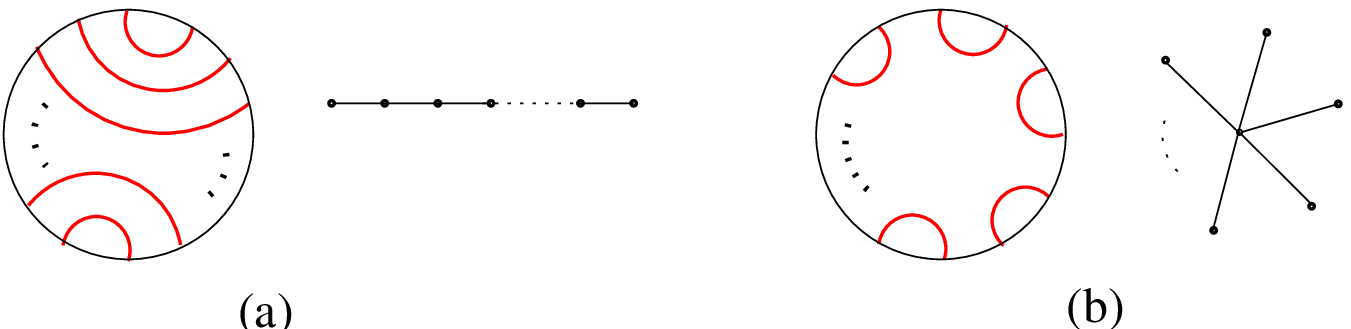}}{10cm}\figlabel\smcomp

This assumes that the eigenvector has been normalised in such a 
way that its smallest component be equal to 1. 
This smallest component corresponds to the link 
patterns shown on figure \smcomp(a), with $n$ {\it nested } arches, 
or in the alternative dual picture, 
to linear trees, and it is possible to prove, independently of
Conjecture 1,  that there is a unique FPL configuration for each
such link pattern  \NAM. 

Then, another conjecture deals with the {\it largest}  component:\par
\noindent{\bf Conjecture 2.} \BdGN\ {\sl The largest component of the
eigenvector occurs for link patterns of $n$ level 1 arches, see 
figure \smcomp(b), and equals $A_{n-1}$, i.e. the total number of
FPL (or ASM) of size 
$n-1$}.

In the present work, we have taken Conjecture 1 for granted
and used the linear problem to compute the $A_n(\pi)$
up to $n=11$. We have found helpful to use 
the symmetry properties of sect. 1 to reduce the dimension of the 
problem. The Hamiltonian $H$ commutes with 
the generators of the group $D_n$ and  the eigenvector of 
largest eigenvalue is
expected to be completely symmetric under these symmetries, in
agreement with Conjecture 1 and Wieland's theorem.
One may thus determine the $A_n(\pi)$ by looking at a reduced 
Hamiltonian acting on  orbits. As  a glance at Table 1 above will
convince the reader, this results in a large gain of computing time
and size. In practice, we have been able to determine all the 
$A_n(\pi)$ up to $n=11$ with an unsophisticated Mathematica code. 
The following conjectures have been extracted from the analysis 
and extrapolation of these data (which are available on request).

\newsec{New results and conjectures}

\penalty 10000
\subsec{Expression of $A_n(\pi)$ for several classes of link patterns $\pi$}

\def\Fac{\buildrel{2}\over{!}}
\def\Fac{!^{!}\,}
\def\Fac{\,{\hbox{\tenrm \char'74}\,}}
\def\Fac{{\hbox{\tenrm \char'74}\,}}

\nind
In view of its frequent occurrence, it is convenient to introduce a new 
notation for the ``superfactorial'' 
\eqn\superfac{ 
m\Fac\ := \prod_{r=1}^m r!=\prod_{j=1}^m(m-j+1)^j\ ,\quad (-1)\Fac=0\Fac=1 .}
%

Then all the results up to $n=11$ are consistent with 
\medskip


\eqnn\conjt

\noindent{\bf Conjecture 3.}\quad {\epsfxsize=18mm\hbox{\raise -6mm\hbox{\epsfbox{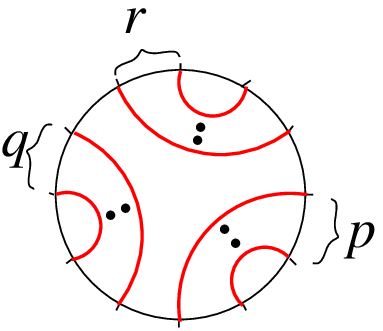}}
$\displaystyle{={(p+q+r-1)\Fac (p-1)\Fac(q-1)\Fac(r-1)\Fac\over
(p+q-1)\Fac 
(q+r-1)\Fac(r+p-1)\Fac}}$\quad $p,q,r,\ge 0$} . \conjt}

\noindent This may also be written in a simpler but less symmetric form, 
using the notation $n=p+q+r$
\eqn\conjtp{
 {{ n-1\choose p}{ n-2\choose p}\cdots{ n-q\choose p}\over
{\rm same \ for \ } n=p+q
} \ . }
But the expert  will also recognize in \conjt\ MacMahon's formula for
plane partitions in a box of size $(p,q,r)$ 
\footnote{$^{(2)}$}{Many thanks to S. Mitra and D. Wilson for this observation.}, i.e.
$$ \prod_{i=1}^p\prod_{j=1}^q\prod_{k=1}^r {i+j+k-1\over i+j+k-2}\ .$$
It would be very interesting to find a bijection between FPL 
configurations with those link patterns and these plane partitions. 

\medskip
\noindent The factorized form does not persist for more complicated
configurations. For example,

\eqnn\conjq

\noindent{\bf Conjecture 4.} \quad
For  $p\ge 1 ,q,r,\ge 0$, \par\noindent
 {\epsfxsize=28mm\hbox{\raise -6mm\hbox{\epsfbox{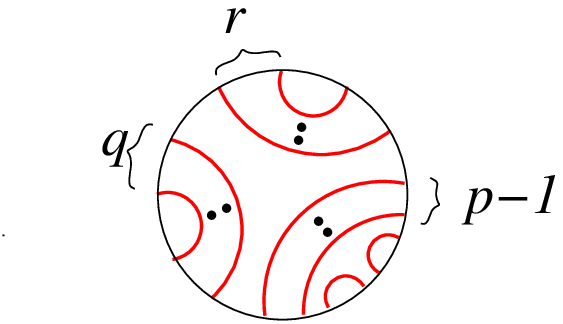}}
$\displaystyle{=
{(q-1)\Fac(r-1)\Fac\over(q+r-1)\Fac }\,
{ p\Fac(p+q+r)\Fac 
\over (p+q+1)\Fac (p+r+1)\Fac}(p+q)!(p+r)!  }\ \ \times
$}\hfill\conjq }\par\noindent
\hbox{\raise -5mm\hbox{$\qquad\qquad\times\ \displaystyle{[p^3+2p^2(q+r+1)+p(q^2+qr+r^2+3(q+r)+1)+q(q+1)+r(r+1)]}$}}

\bigskip
\eqnn\conjc

\noindent{\bf Conjecture 5.} \quad
For  $p\ge 1,q,r,\ge 0$, \par\noindent
 \hglue-10mm{\epsfxsize=23mm\hbox{\raise
-6mm\hbox{\epsfbox{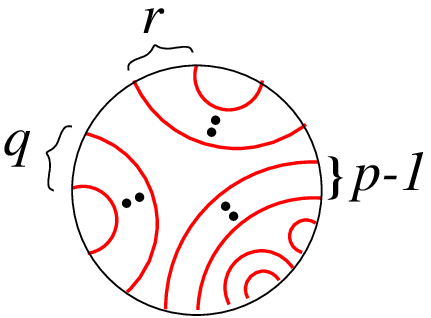}}
$\!\!\!\! 
=\displaystyle
{(q-1)\Fac(r-1)\Fac\over 2  (q+r-1)\Fac }
{ (p+1)\Fac(p+q+r+1)\Fac 
\over (p+q+3)\Fac (p+r+3)\Fac} \ (p+q+2)!(p+q+1)!(p+r+3)!(p+r)!$}}
\par\noindent $\quad\times
(p+2) \Big[
p^5 + p^4 (7  + 4\,q + 4\,r )
+ p^3 (17 + 22\,q + 6\,q^2  + 24\,r + 10\,q\,r +   6\,r^2 )$ 
\par\noindent $ \qquad
+ p^2(17  + 40\,q + 24\,q^2 + 4\,q^3 + 46\,r + 42\,q\,r 
+ 8\,q^2\,r  + 30\,r^2 +   8\,q\,r^2 + 4\,r^3 )$ 
\par\noindent $ \qquad
+ p (6 +  28\,q + 29\,q^2 + 10\,q^3 + q^4 + 32\,r + 49\,q\,r 
+ 17\,q^2\,r +   2\,q^3\,r + 41\,r^2 $ \hfill \conjc
\par\noindent $ \qquad\qquad
+ 23\,q\,r^2 + 3\,q^2\,r^2  + 16\,r^3  +   2\,q\,r^3  +\,r^4)$ 
\par\noindent $ \qquad
+ 6\,q +  11\,q^2 +   6\,q^3 + q^4 +   6\,r 
+   13\,q\,r +   3\,q^2\,r + 15\,r^2 + 15\,q\,r^2 
+ 3\,q^2\,r^2 +   12\,r^3 + 2\,q\,r^3
+ 3\,r^4\Big]$


\subsec{Polynomial behavior in $n$ and asymptotic behaviour for large $n$ }

\fig{Describing a configuration by a  Dyck path or a
Young diagram}{\epsfbox{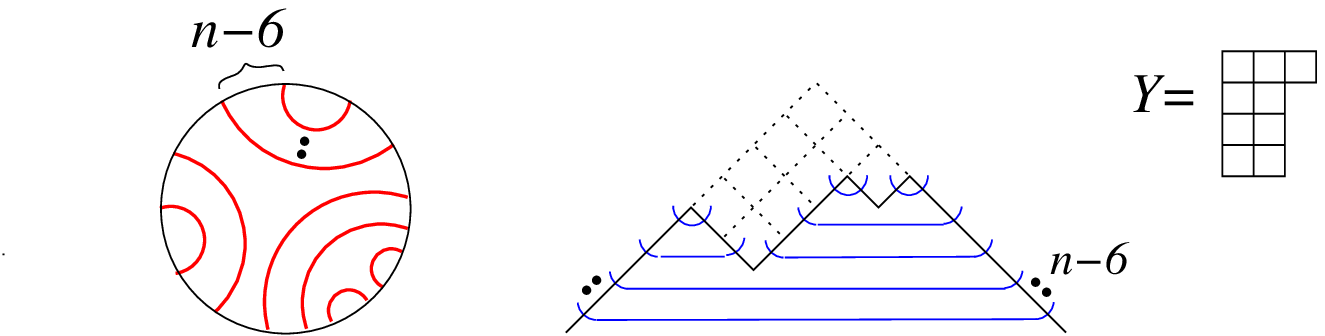}}{9cm}\figlabel\young

\noindent Let us consider link patterns $\pi$ 
made of a given set ${\cal S}$ of $r$ arches plus 
 $n-r$ nested arches as in Conjectures 3 and 4 above, and let
$n$ vary, while keeping ${\cal S}$ fixed. 
Any such link pattern is also encoded by a (Dyck) path, or by the 
complementary Young diagram $Y$, see
Figure \young\footnote{$^{(3)}$}{The ambiguity between
the Young diagram $Y$ and its transpose in this definition will be 
immaterial in what follows.}. 
We denote by $|Y|$ the number of boxes of $Y$ and
by $\dim Y$ the dimension of the representation of the 
symmetric group $S_{|Y|}$ labelled by $Y$. We recall (see for example 
\Stan)
 the useful expression for the ratio $ {{\rm
dim\,}Y\over|Y|!}={1\over {\rm hl}(Y)}$, the inverse
{\it hook length} of the diagram,  \ie the inverse product of the hook
lengths of all its boxes. 
Finally, we denote by $F(Y)$ the set of diagrams obtained by
adjonction of one box to $Y$ according to the usual
rules. Alternatively, if $D_Y$ is the corresponding
 irreducible representation of $Sl(N)$,
$F(Y)$ labels the set of representations appearing in the
decomposition into irreducibles  of $D_{\copy101}\otimes D_Y$. Then

\noindent{\bf Conjecture 6.} {\sl For $n\ge r$ 
\eqn\polyf{ A_n(\pi)={1\over  |Y|!} P_{Y}(n) }
where $P_{Y}(n) $ is a polynomial of degree $|Y|$ with coefficients
in $\Bbb{Z}$ and its highest degree coefficient is equal to $\dim Y$.}

\noindent For example, in the case covered by equation \conjtp, 
$Y$ is a rectangular $p\times q$ Young diagram,  $|Y|= p q$
and $ (p q)! { 2! \cdots (q-1)!\over p! (p+1)! \cdots (p+q-1)!}$ is
indeed an integer. See more examples in Appendix A.
\ommit{ Another example is provided by\ \  
 {\epsfxsize=20mm\hbox{\raise -8mm\hbox{\epsfbox{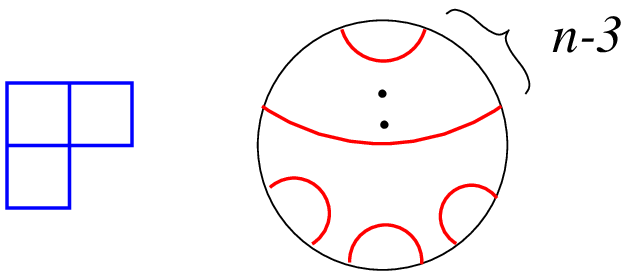}}
$={1\over 6} (n-2)(2n^2-5n+9)$}}.}

\fig{Configuration described by
 two Young diagrams}{\epsfbox{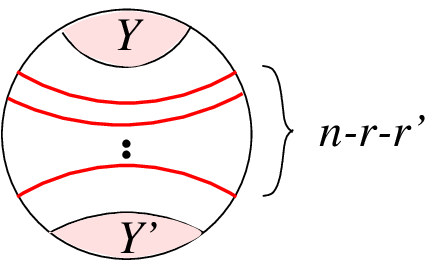}}{3cm}\figlabel\youngd

As a corollary of Conjecture 6,  
the asymptotic behavior for large $n$ is given  by
\eqn\asymp{ A_n(\pi)  \approx {\dim Y\over|Y|!}n^{|Y|}\ .}
Such an asymptotic behavior had been observed in the case 
of open boundary
conditions by Di Francesco \PDF, who derived it as a consequence 
of the eigenvector equation. 
The action of the Temperley-Lieb generator $e_i$ on an open link pattern
associated with one Young diagram $Y$ or on the corresponding Dyck path
is described by the ``raise and peel'' process of \dGNPR:
the resulting Young diagram $\bY$ is either $Y$ itself if the site $i$ 
is a local peak of the path, or has one less box than $Y$ if
$i$ is a local minimum of the path (and then $Y\in F(\bY)$), 
or is a diagram with a larger 
number of boxes than $Y$ otherwise.  What changes in the case of periodic 
boundary conditions is the possibility of an action on the ``other 
side'' of the link pattern. In order to carry out the discussion 
in the periodic case,   we thus have to generalize our
considerations to  configurations 
described by {\it two} Young diagrams $Y$ and $Y'$, 
with $r$ and $r'$ arches,  separated by a  number $n-r-r'$ of
parallel arches (see Fig. \youngd ). Then

\noindent{\bf Conjecture 7.} {\sl For $n\ge r+r'$ 
\eqn\conjse{ A_n(\pi)=:A_n(Y,Y')={1\over |Y|!  |Y'|!} P_{Y,Y'}(n) }
with  $P_{Y,Y'}(n) $  a polynomial of degree $|Y|+|Y'|$ 
with coefficients in $\Bbb{Z}$ and its highest degree 
coefficient is $\dim Y \, \dim Y'$.}

\noindent This is exemplified on the configurations of  Conjectures 4 or
5:  for given $q$ and $r$, one Young diagram is a $q\times r$
rectangle, the other is made of one or two boxes, and $Y$ and $Y'$
are separated by $p-1$ arches; then  in the expressions given 
in Conj. 4 or 5, 
the first factor represents ${\dim Y\over  |Y|!}
{\dim Y'\over  |Y'|!}$, the second (the ratio of superfactorials)
is  seen to be a polynomial in $p$, and the degree of the whole 
expression is easily computed. 

Again, one derives from this conjecture  the asymptotic behavior 
\eqn\asympt{ A_n(Y,Y') \approx {\dim Y\over  |Y|!}
{\dim Y'\over  |Y'|!} n^{ |Y|+ |Y'|}}
We shall now show  that this asymptotic behavior  is 
consistent with the eigenvector equation \eigenv. 
\ommit{} 
{Let $\pi_a$ be a link pattern described by a pair of
Young diagrams $(Y,Y')$, as in Fig. \youngd, and
$e_i$ be a generator of the periodic
Temperley-Lieb algebra. The link pattern  $\pi_b=e_i \pi_a$
is described by a pair $(\bY,\bY')$.}
Identifying the coefficient of $\pi_b$ in \eigenv\ and 
using the Ansatz \asympt, we 
find that for $n$ large, the only terms to contribute are 
either $Y=\bY$, $Y'\in F(\bY')$ or  $Y\in F(\bY)$, $Y'=\bY'$
\eqn\leading{
2 n A_n(\bY,\bY')=\sum_{Y\in F(\bY)} A_n(Y,\bY')
+ \sum_{Y'\in F(\bY')} A_n(\bY, Y') +O({1\over n})}
which is consistent with the behaviour \asympt, since 
$$ 2 {\dim \bY\over |\bY|! }{\dim \bY'\over |\bY'|! }
=\sum_{Y\in F(\bY) } {\dim Y\over |Y|! }{\dim \bY'\over
|\bY'|! }
+\sum_{Y'\in F(\bY') }{\dim \bY\over |\bY|! } {\dim Y'\over
|Y'|! }$$
which results itself from the identity
\eqn\identity{ {\dim \bY\over |\bY|! }=\sum_{Y\in F(\bY) }
{\dim Y\over |Y|! }\ .}

\ommit{\noindent{\bf Conjecture 7}
{\sl The leading large $n$ behaviour is given by}
$$ A_n(\pi)= P_{Y}(n) \approx {{\rm
dim\,}Y\over|Y|!}n^{|Y|}$$ 
Such expressions had been observed before by Di Francesco in the 
case of open boundary conditions, which, in the large $n$ limit, 
should not differ from the periodic case under study~\PDF.}

\subsec{Recursion formulae generalizing Conjecture 2}

\noindent 
In the same way as Conjecture 2 relates the number of FPL
configurations for a certain link pattern, made of $n$ 
simple arches, to the {\it inclusive sum} of all FPL configurations 
of size $n-1$, one finds relations between other 
configuration numbers of size $n$ and inclusive sums of size  $n-1$.

\fig{Relating FPL configurations of size $n$ with inclusive
configurations of size $n-1$}{\epsfbox{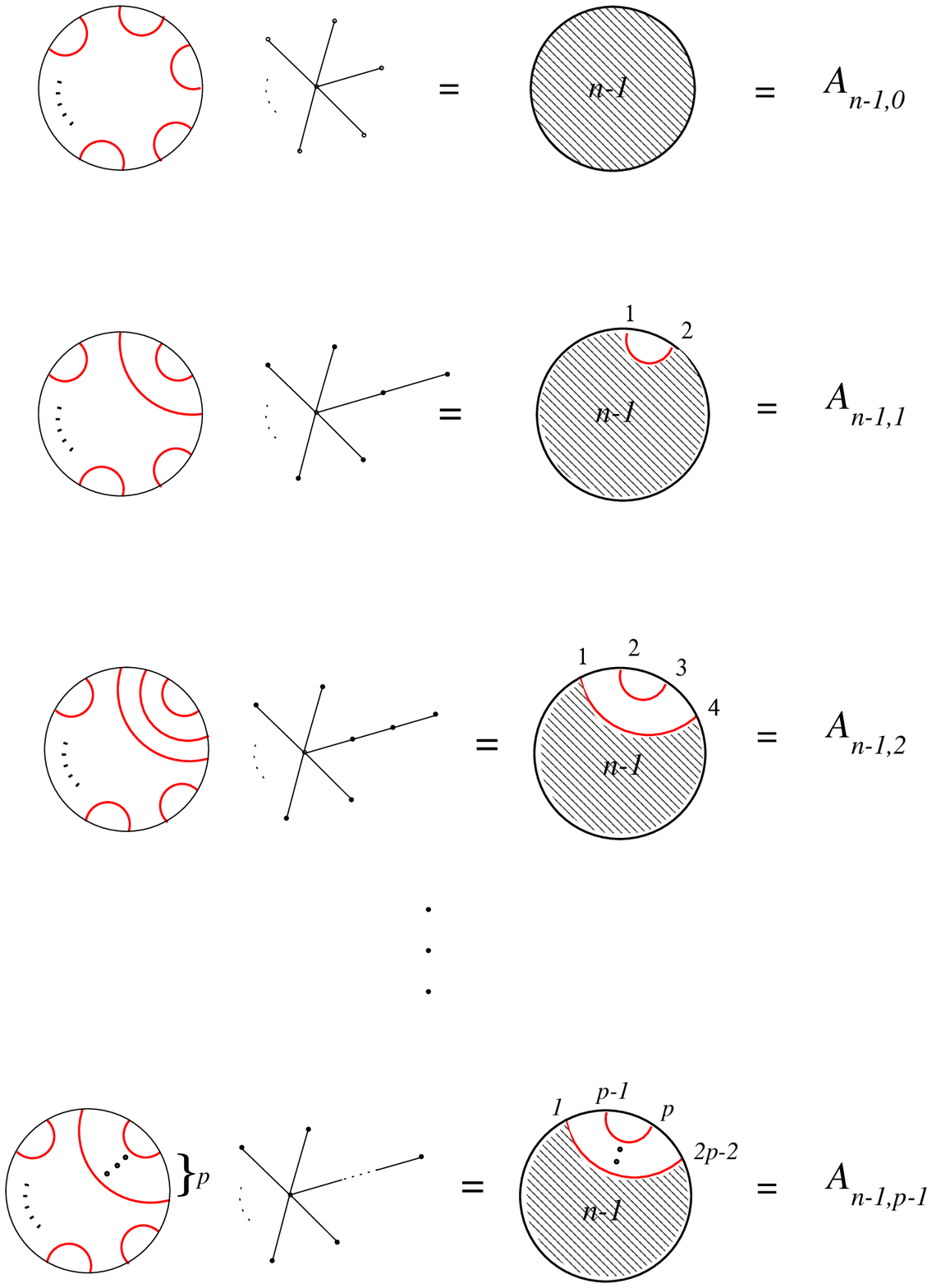}}{9cm}\figlabel\newid
\noindent{\bf Conjecture 8.}
(i) \Wie\ {\sl We have the relations depicted on Figure
\newid, where for example the expression $A_{n-1,1}$ 
on the r.h.s. is the  number of FPL configurations of
 size $n-1$ containing an arch between external links 1 and 2.}  \par 
(ii) {\sl The rhs of these relations,  
{\rm at size $n$}, take  respectively the values}

$$
A_{n,0}=A_n\,,\ \ 
A_{n,1}= {3\over 2}{n^2+1\over (2n-1)(2n+1)} A_n\,,\ \  
A_{n,2}= {1\over 16}{59 n^6+299n^4+866n^2+576\over
(2n-3)(2n-1)^2(2n+1)^2(2n+3)}A_n$$ and $$\!\!\!\!\!\! 
A_{n,3}={3\over 512}
{2579\, n^{12}+     39364\, n^{10}  + 374412\, n^8 
 + 2174092\ n^6 + 6601109\, n^4 + 11674044\, n^2 +6350400
\over(2n-5)(2n-3)^2(2n-1)^3(2n+1)^3(2n+3)^2(2n+5) }A_n 
$$ 
It is easy to guess the 
general form of 
$A_{n,p}=( P_{p(p+1)}(n^2)/ \prod_{\ell=1}^{p}
  (4n^2-(2\ell-1)^2)^{p+1-\ell})  A_n$
as a ratio of two even 
polynomials of degree $p(p+1)$ in $n$, 
although the detailed form of the numerator remains unclear.
The  expressions of $A_{n,p}$, $p=1,2$ in (ii) were known to D. Wilson
\DW, while
the one of  $A_{n,3}$ seems to be new.

\fig{Relating FPL configurations of size $n$ with inclusive
configurations of size $n-1$, cont'd}
{\epsfbox{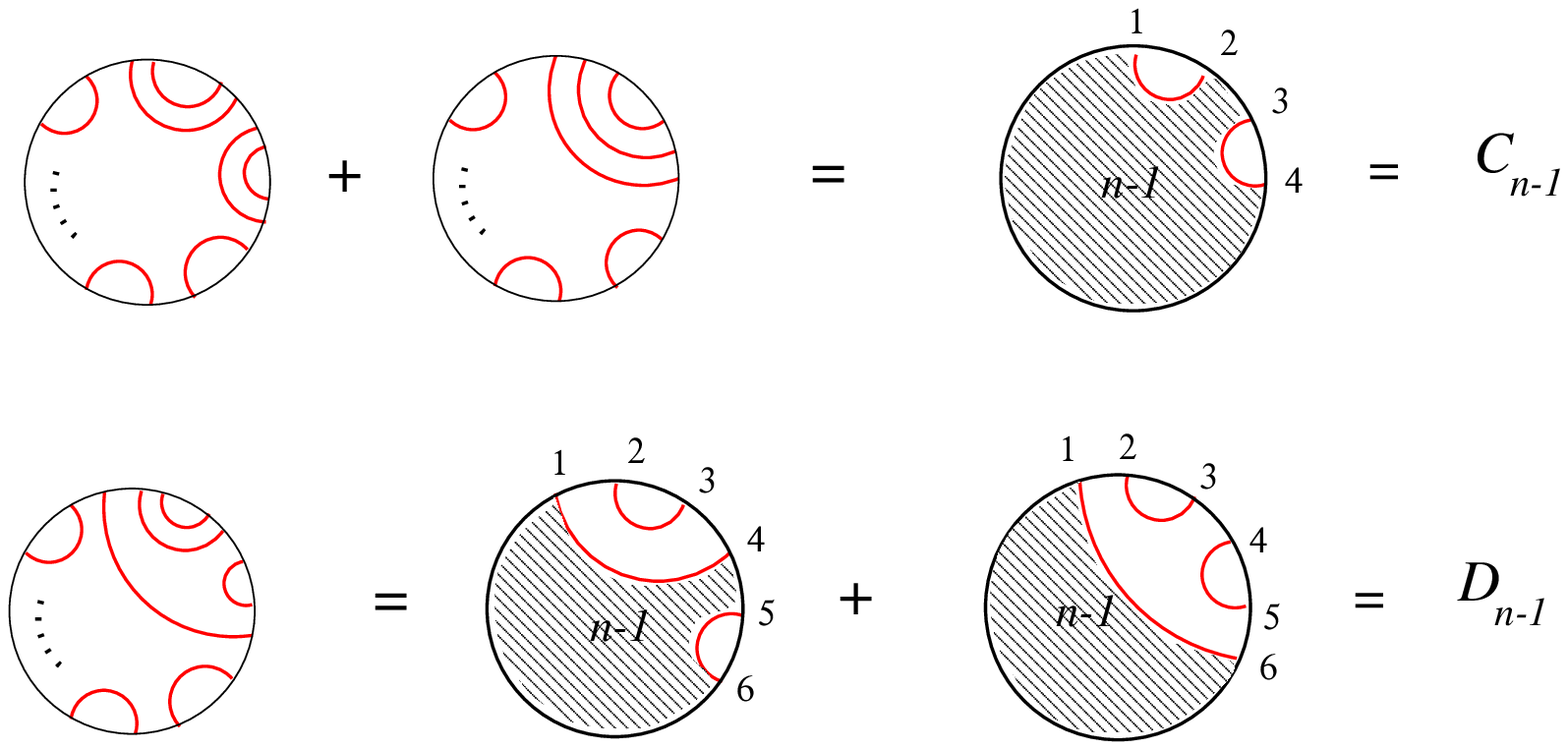}}{9cm}\figlabel\newidd
\noindent{\bf Conjecture 9.} {\sl There are equalities 
as shown on  Figure \newidd\
between  the  {\it sum} of two configuration numbers $A_n(\pi)$ 
and an inclusive sum $C_{n-1}$ of size $n-1$, or vice versa,  with}
$$\eqalignno{ 
 C_n& = {{97\,n^6} + {82\,n^4} 
- {107\,n^2}  -792 \over 8 (2n-3)(2n-1)^2(2n+1)^2(2n+3)}A_n\cr
D_n&={9\over 256} {  5977\,n^{12}+ 16622\,n^{10}+ 54681\,n^8-
216784\,n^6 - 2071808\,n^4- 337488\,n^2+3456000 \over 
(2n-5)(2n-3)^2(2n-1)^3(2n+1)^3(2n+3)^2(2n+5) }A_n\ . \cr}
$$
By combining the previous formulae it follows that for $n\ge 3$
\medskip 
\centerline{\epsfxsize=10mm\hbox{\raise
-4mm\hbox{\epsfbox{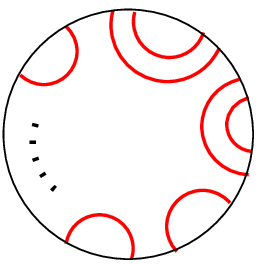}} $\Bigg|_{n+1} \displaystyle{= {3^3.5\over
2^4}{(n^2-4)(n^4+3n^2+4)\over
 (2n-3)(2n-1)^2(2n+1)^2(2n+3)}A_n}\ . $}}

\medskip
These identities are just the beginning of a host of
relations, such as
\vglue3mm
\epsfxsize=130mm{\epsfbox{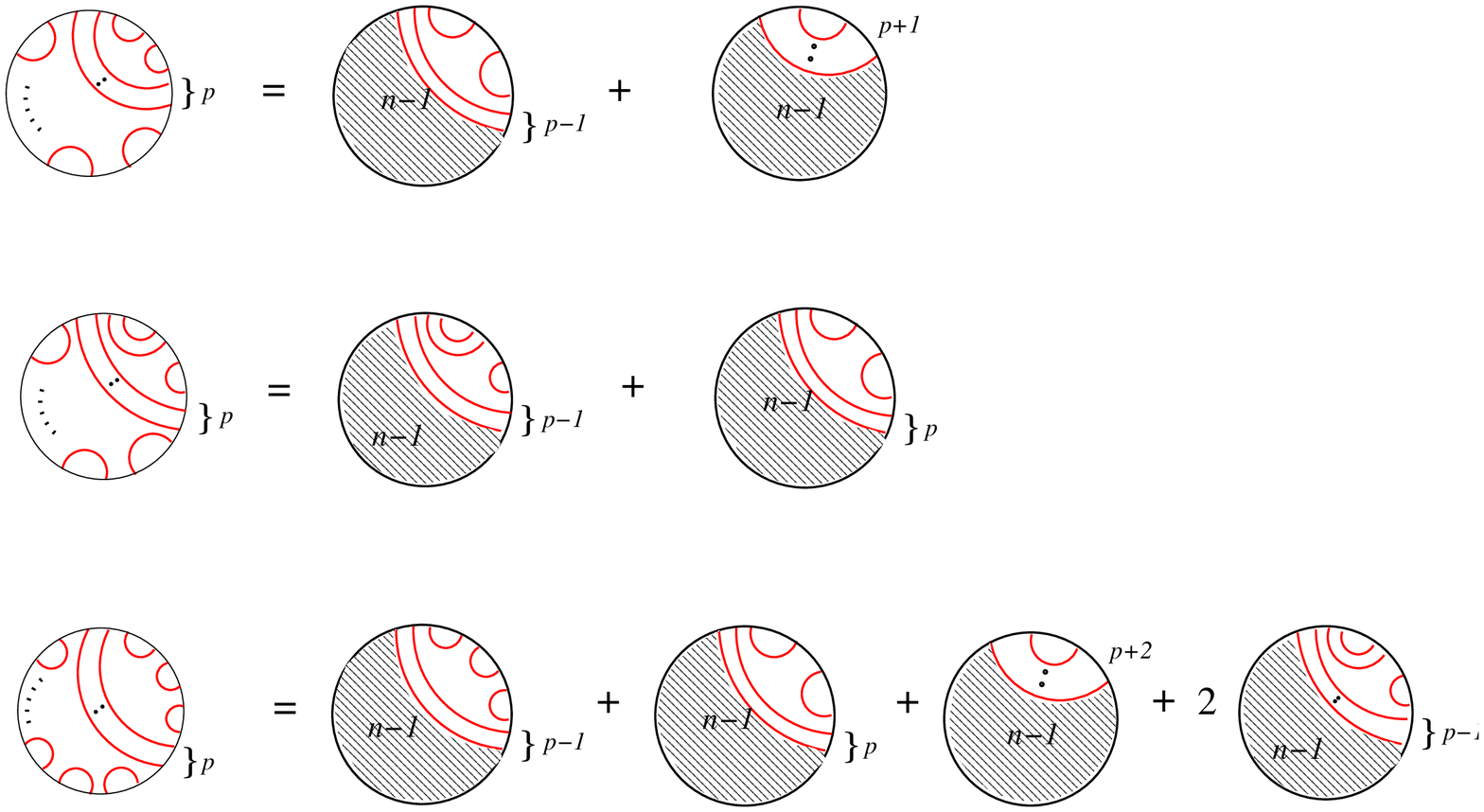}}

 \noindent but their systematics has remained elusive so far.


One may also conjecture that
{\epsfxsize=10mm\hbox{\raise
-4mm\hbox{\epsfbox{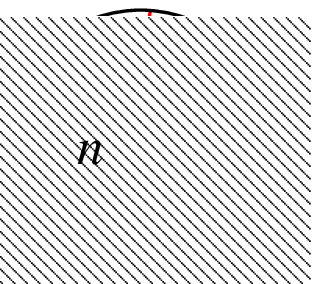}}}}  and 
{\epsfxsize=10mm\hbox{\raise
-4mm\hbox{\epsfbox{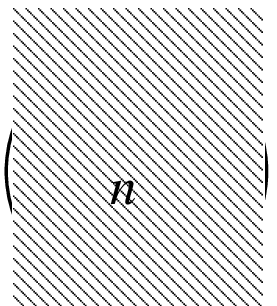}}}} are again both of the form $$P_{12}(n)
A_n/\((4n^2-25)(4n^2-9)^2(4n^2-1)^3\)$$  with the even polynomials $P_{12}(n)$
equal to respectively 
$$
\scriptstyle{{3\over 512}\(
12631\, n^{12} + 101096\, n^{10} + 586518\, n^8 
+ 1237988\, n^6 - 5800349\, n^4  - 19336284\, n^2 -23976000
\)}$$
and
$$\scriptstyle{{3\over 512}\(
 23231\, n^{12} - 1364\, n^{10} - 258432\, n^8  - 2538692\, n^6 
- 6630499\, n^4 + 17311356\, n^2 + 44712000 
\)}\ .$$
The expression of $A_{n,3}+{\epsfxsize=8mm\hbox{\raise
-3mm\hbox{\epsfbox{newid4.eps}}}}$ was known to D. Wilson \DW.


\newsec{Discussion}
\noindent
This paper has presented a certain number of
conjectural expressions and recursion formulae for
the numbers of configurations of FPL with periodic boundary
conditions. At this stage all these expressions remain empirical, 
and based on the actual data of the linear problem. The connection 
with the numbers of FPL thus relies on another conjecture (Conjecture 1). 
In some cases, however, the numbers given in this paper have been 
tested against the direct computation of FPL configurations \NAM. 
A similar discussion is currently being carried out for the other types
of boundary conditions by another group \MNdGB.

More conjectural expressions have been collected for  other
 types of configurations (see Appendix A), but this seems a gratuitous game
 in the absence of a guiding principle.
Observe however the simplicity of the ``three-point-functions'' (Conjecture 3)
as compared to the cumbersomeness  of the others. Could 
 this suggest that the latter may be obtained from the former, 
in the same way as higher correlation functions in Conformal Field
Theories, say, may be constructed 
from the 3-point functions~?



{\vskip1cm
\centerline{\bf Acknowledgments}
It is a pleasure to acknowledge fruitful discussions with
Philippe Di Francesco, 
Saibal Mitra, Nguyen Anh Minh and Paul Pearce
and to thank  Bernard Nienhuis for communication of the 
manuscript of \MNdGB\ prior to publication, and 
David Wilson for discussions and for suggestions on this
manuscript. 
This work is partially supported by the European network 
HPRN-CT-2002-00325. }

\ommit{
\eqn\truc{}
\eqna\Ib
$$\eqalignno{
Z&= Z_1   &\Ib a\cr
F &= F\big(1\big)&\Ib b\cr
}$$
$$\eqalign{G_4(g)&=2 + 10\ g + 74\ g^2 + 706\ g^3 + 8162\ g^4 + 110410\ g^5 + 
    1708394\ g^6 \cr &\qquad+ 29752066\ g^7 + 576037442\ g^8 + 12277827850\
g^9 +285764591114\ g^{10}+O(g^{11})}$$
\eqnn\Fzer
$$\eqalignno{F^{(0)}(g):=F^{(0)}(g,1)&=\log a^2-{1\over
12}(a^2-1)(9-a^2)\cr
&=2\sum_{n=1}^\infty  (3g)^n  {(2n-1)!!\over n!(n+2)!}\cr}$$
}


\listrefs


\appendix{A}{More configurations } 

 {\epsfxsize=30mm\hbox{\raise -6mm\hbox{\epsfbox{otherex.eps}}
$={(n-2)\over 6}(2n^2-5n+9)$}}

\bigskip
 {\epsfxsize=30mm\hbox{\raise -6mm\hbox{\epsfbox{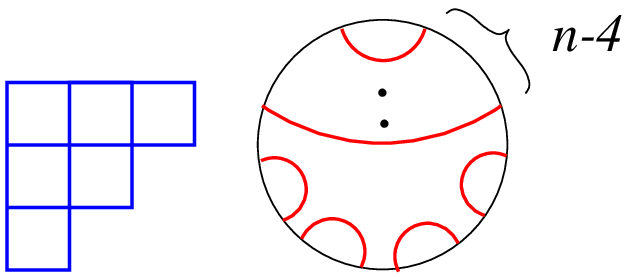}}
$={(n-1)(n-3)\over 180} (4n^4-32n^3+155n^2-394n+540)$}}

\bigskip
 {\epsfxsize=30mm\hbox{\raise -6mm\hbox{\epsfbox{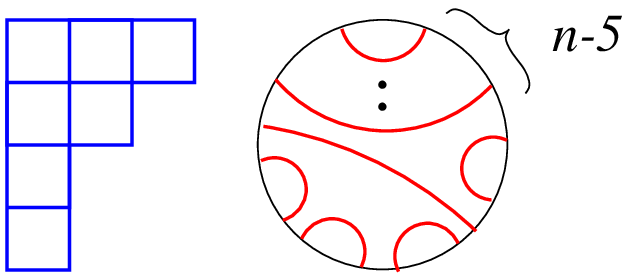}}
$={(n-1)(n-3)(n-4)\over 720}(5n^4-38n^3+197n^2-522n+840)$}}

\bigskip
 {\epsfxsize=32mm\hbox{\raise -6mm\hbox{\epsfbox{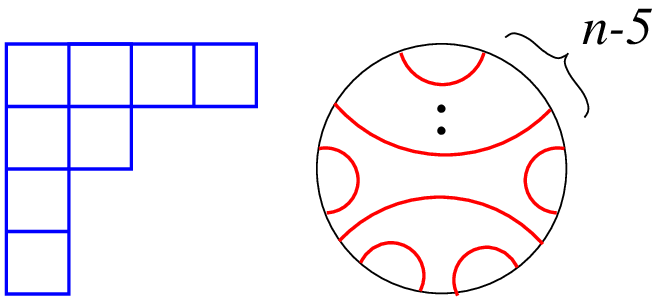}}
$={(n-1)(n-4)\over 20160}
\scriptstyle{(45n^6-635n^5+4639n^4-21865n^3+68924n^2-136740n+146160)}$}}
\bigskip
 {\epsfxsize=30mm\hbox{\raise -6mm\hbox{\epsfbox{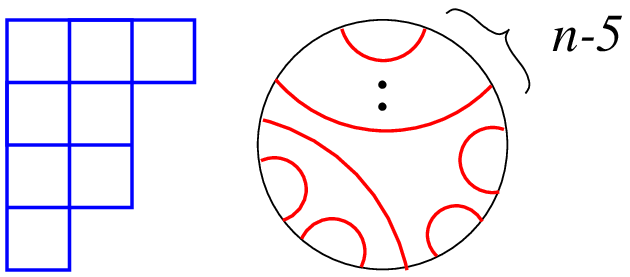}}
$={(n-1)(n-2)(n-3)(n-4)\over 4!5!}(5n^4-46n^3+275n^2-802n+1440)$}}

\bigskip
 {\epsfxsize=32mm\hbox{\raise -6mm\hbox{\epsfbox{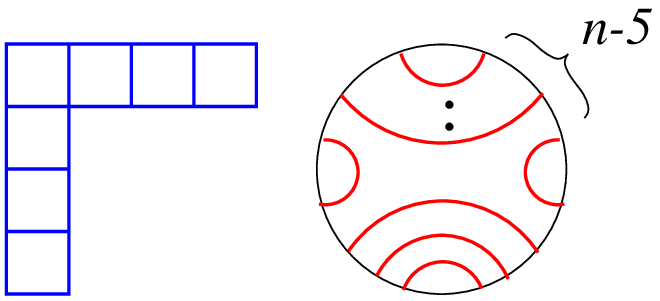}}
$={(n-4)\over 2520}(10n^6-135 n^5+853n^4-3378n^3+9343n^2-17403n+18270)$}}

\bye